\documentclass[12pt]{article}
\usepackage{amsmath}
\usepackage{bm}
\usepackage{graphicx}
\begin{document}
%\baselineskip 20pt \setlength\tabcolsep{2.5mm}
%\renewcommand\arraystretch{1.1}
%\setlength{\abovecaptionskip}{0.1cm}
%\setlength{\belowcaptionskip}{0.5cm}
%%%%%%%%%%%%%%%%%
\title { The study of participant-spectator matter and collision
dynamics in heavy-ion collisions}
%%%%%%%%%%%%%%%%%%%%%%%%%%%%%%%%
\author {Aman D. Sood and  Rajeev K. Puri\\
\it Department of Physics, Panjab University, Chandigarh -160 014,
India.\\} \maketitle
\begin{abstract}
We present the simulations of heavy-ion collisions in terms of
participant-spectator matter. We find that this matter depends
crucially on the collision dynamics and history of the nucleons.
The important changes in the momentum space are due to the binary
nucleon-nucleon collisions experienced during the high dense
phase. This was otherwise not possible with mean field alone. The
collisions push the colliding nucleons into midrapidity region
responsible for the formation of participant matter. This
ultimately leads to thermalization in heavy-ion collisions.
\end{abstract}
%PACS number: 25.70.-z, 25.70.Jj \\
 Electronic address:~rkpuri@pu.ac.in
\newpage
\section{Introduction}
The heavy-ion collisions at intermediate energies are the center
of present day nuclear research. This is because of several rare
phenomena emerging at these energies and also their utility in
several other branches of physics.\cite{gupt88,aich91,stoc86} The
study of hot and dense matter and its relation to nuclear equation
of state and cross section has always fascinated the
researchers.\cite{gupt88,aich91,stoc86,hartphd,blat91,mage0061}
Some of the rare phenomena emerging are the
multifragmentation,\cite{gupt88,aich91,stoc86} collective
flow,\cite{aich91,stoc86} stopping\cite{aich91,stoc86} as well as
subthreshold particle production\cite{grei99} etc. In the low
energy heavy-ion collisions, the fusion and decay of excited
compound nucleus and fission dominate the physics.\cite{puri91}
Whereas at high incident energies, the transparency and complete
disassembly of nuclear matter happens. In terms of theoretical
description, the relative dominance of real and imaginary parts of
G-matrix decides the fate of a reaction.\cite{hart98} At low
incident energies, the Pauli principle forbids the nucleon-nucleon
binary collisions and scattering.\cite{gupt88,aich91,stoc86} The
attractive mean field decides the fate of colliding pairs. This
results into fusion and consequently, into its decay. The frequent
nucleon-nucleon collisions at high incident energies make
imaginary part very significant. However, both real and imaginary
parts of complex G-matrix are equally important at intermediate
energies. This picture can also be looked in terms of
participant-spectator matter and fireball
concept.\cite{gupt88,sood04} The nucleon-nucleon collisions play
an important role in destroying the mutual initial nucleon-nucleon
correlations and memory of nucleons. This ultimately leads to the
above rare phenomena at intermediate energies.

We here plan to understand the formation of participant-spectator
matter and its development in terms of collision dynamics.
Recently, participant-spectator matter has been found to be
independent of the system size at energy of vanishing flow.
Therefore, it acts as a barometer for the study of disappearance
of flow and balance energy.\cite{sood04,sood04a} Section 2,
describes the model briefly. In section 3, we present our results
and section 4 summarizes the outcome.
%%%%%%%%%%%%%%%%%%%%%%%%%%%%%%%%5

\section{The model}

We simulate the nucleons within the framework of quantum molecular
dynamics (QMD) model. In the QMD model,\cite{aich91,stoc86} each
nucleon propagates under the influence of mutual interactions. The
propagation is governed by the classical equations of motion:
%%%%%%%%%%%%%%%%%%%%%%%%%%%%%%%%%%
\begin{equation}
\dot{{\bf r}}_i~=~\frac{\partial H}{\partial{\bf p}_i}; ~\dot{{\bf
p}}_i~=~-\frac{\partial H}{\partial{\bf r}_i},
\end{equation}
%%%%%%%%%%%%%%%%%%%%%%%%%555
where \emph{H} stands for the Hamiltonian which is given by:
%%%%%%%%%%%%%%%%%%%%%%%%%%%%%%%%%%%%%%5
\begin{equation}
H = \sum_i^{A} {\frac{{\bf p}_i^2}{2m_i}} + \sum_i^{A}
({V_i^{Skyrme} + V_i^{Yuk} + V_i^{Coul}}).
\end{equation}
%%%%%%%%%%%%%%%%%%%%%%%%%%%%%%
The $V_{i}^{Skyrme}$, $V_{i}^{Yuk}$, and $V_{i}^{Coul}$ in Eq. (2)
are, respectively, the Skyrme, Yukawa, and Coulomb potentials.

%%%%%%%%%%%%%%%%%%%%%%%%%%%%%%%%%%%%%%%%%%%%5

\section{Results and Discussion}

We here simulated the reactions of $^{40}$Ca+$^{40}$Ca and
$^{131}$Xe+$^{131}$Xe, at different colliding geometries. This
varies from very central ($b=0$ fm) to peripheral one
($b=R_{1}+R_{2}$; $R_{i}$ is the radius of either target or
projectile). The incident energy is also varied between 200 and
400 MeV/nucleon. In the present study, we use a hard equation of
state along with nucleon-nucleon cross section $\sigma=40$ mb. As
discussed by many authors, the dynamics is insensitive towards the
nuclear equation of state as well as towards nucleon-nucleon cross
section. However, the above equation of state and cross section
has been reported to reproduce the balance energy over wide range
of colliding masses.\cite{sood04a}

The present participant-spectator matter demonstration is based on
the definition reported in our earlier recent
publication.\cite{sood04a} Here participant-spectator matter is
defined in two different ways. (a) In the first definition, all
nucleons experiencing at least one collision are labeled as
participant matter. The remaining nucleons are the part of
spectator matter. (b) In the second definition, the above
demarcation is based on the experimental method, where one uses
different rapidity cuts to define participant and spectator
matter. The rapidity of \emph{i}th particle is defined as
\begin{equation}
Y(i) = \frac{1}{2}\ln\frac {{\bf{E}}(i)+{\bf{p}}_{z}(i)}
{{\bf{E}}(i)-{\bf{p}}_{z}(i)},
\end{equation}
where ${\bf E}(i)$ and ${\bf p}_{z}(i)$ are, respectively, the
total energy and longitudinal momentum of {\it i}th particle. Now
one can impose different cuts to study the different
participant-spectator matter. We shall here use the first
definition to construct the participant-spectator matter. However,
one should note that both these definitions have been reported to
give same results.\cite{sood04}

\begin{figure}[!t] \centering
 \vskip 0.5cm
\includegraphics[angle=0,width=10cm]{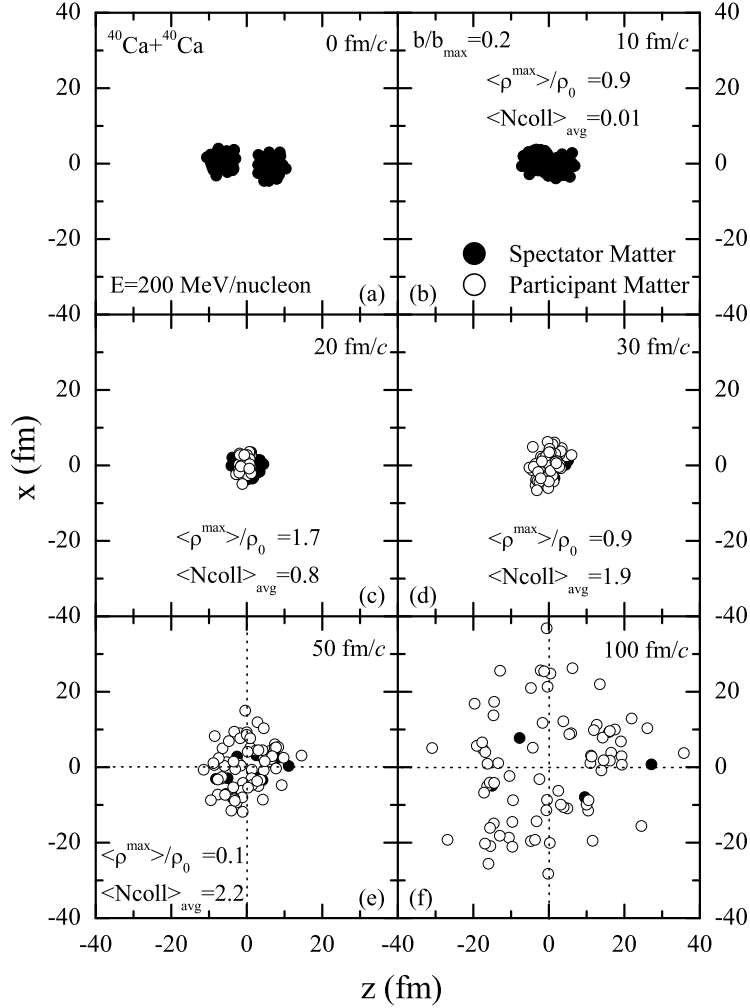}
\vskip -0.2 cm \caption{The time evolution of a single event of
$^{40}$Ca+$^{40}$Ca reaction at 200 MeV/nucleon and
$b/b_{max}=0.2$ in coordinate space. The solid (open) circles
represent spectator (participant) matter, respectively. The
$<\rho^{max}>/\rho_{0}$ and $<Ncoll>_{avg}$ are the maximum
density and average number of nucleon-nucleon collisions averaged
over large number of events.}\label{6.1p}
\end{figure}

\begin{figure}[!t] \centering
 \vskip 0.5cm
\includegraphics[angle=0,width=10cm]{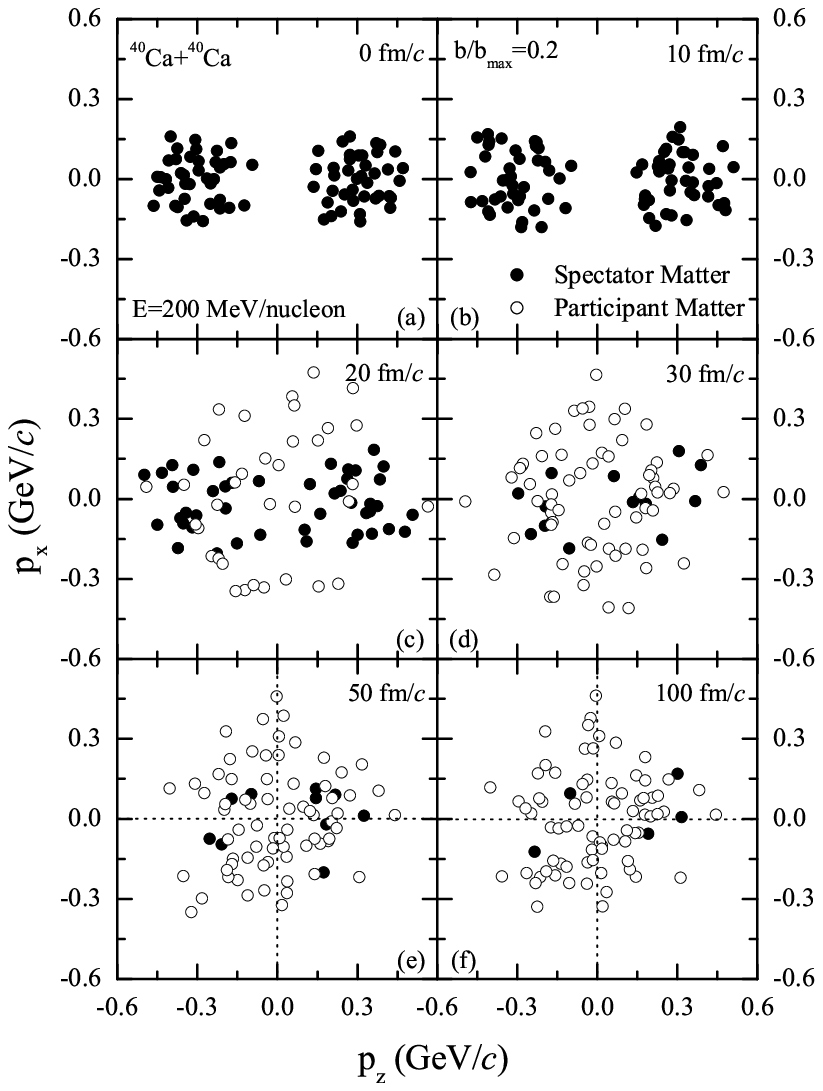}
\vskip -0.2 cm \caption{Same as Fig. \ref{6.1p}, but in momentum
space.}\label{6.2p}
\end{figure}

In Figs. \ref{6.1p} and \ref{6.2p}, we display the time evolution,
respectively, in the spatial and momentum spaces for the central
collision of $^{40}$Ca+$^{40}$Ca at 200 MeV/nucleon. The solid
(open) circles denote the spectator (participant) matter. During
initial stages of the reaction (till 10 fm/{\it c}), projectile
and target are well separated both in the coordinate and momentum
spaces. As a result, whole of the matter is in the form of
spectator matter only. As the nuclei overlap, the density
increases, that leads to more and more nucleon-nucleon binary
collisions. This changes the matter from spectator to participant
one. Due to the frequent nucleon-nucleon collisions, nucleons
scatter in the transverse direction also. With the evolution of
the reaction, participant matter increases, leading to more
nucleons scattering in the transverse direction.
%%%%%%%%%%%%%%%%%%%%%%%%%%%%%%%%%%%%%%%%%%%%%%%%%%%%%%%%%%%%%%%%%%%%%%%%%%%%%%%%%%
%%%%%%%%%%%%%%%%%%%%%%%%%%%%%%%%%%%%%%%%%%%%%%%%%%%%%%%%%%%%%%%%%%%%%%%%%%%%%%%%%%
At the end of the reaction, we see only a few spectator nucleons.

Looking at the momentum space (Fig. \ref{6.2p}), we see that the
spectator nucleons have very little change in their initial
velocity profile over the period of the reaction. On the other
hand, participant matter leads the matter into midrapidity region.
This drives the system into global equilibrium. Naturally, more
the initial correlations are destroyed, more is the system close
to the global equilibrium. Since, these figures are the snapshots
of a single event, the picture could be different for different
events. The matter is cold and fragmented at the final state. This
may lead to the idea that compressed matter expands and due to
Coulomb instabilities, this breaks into pieces. However, this
argument has been questioned by one of us and collaborators who
were able to detect the fragments at the time when matter is
compressed and hot.\cite{goss97} If one evolutes the reaction
between time steps 10-30 fm/\emph{c}, one sees that the density
rises from 0.9 to 1.7 between 10-20 fm/\emph{c}, and falls by the
same amount between 20-30 fm/\emph{c}. However, the average
collision number is not symmetric. The collision rate between
20-30 fm/\emph{c} is higher than between 10-20 fm/\emph{c}. This
points toward the compactness of the nucleons in space with slower
velocities.
%%%%%%%%%%%%%%%%%%%%%%%%%%%%%%%%%%%%%%%%%%%%%%%%%%%%%%%%%%%%%%%%%%%%%%%

\begin{figure}[!t] \centering
 \vskip 0.5cm
\includegraphics[angle=0,width=10cm]{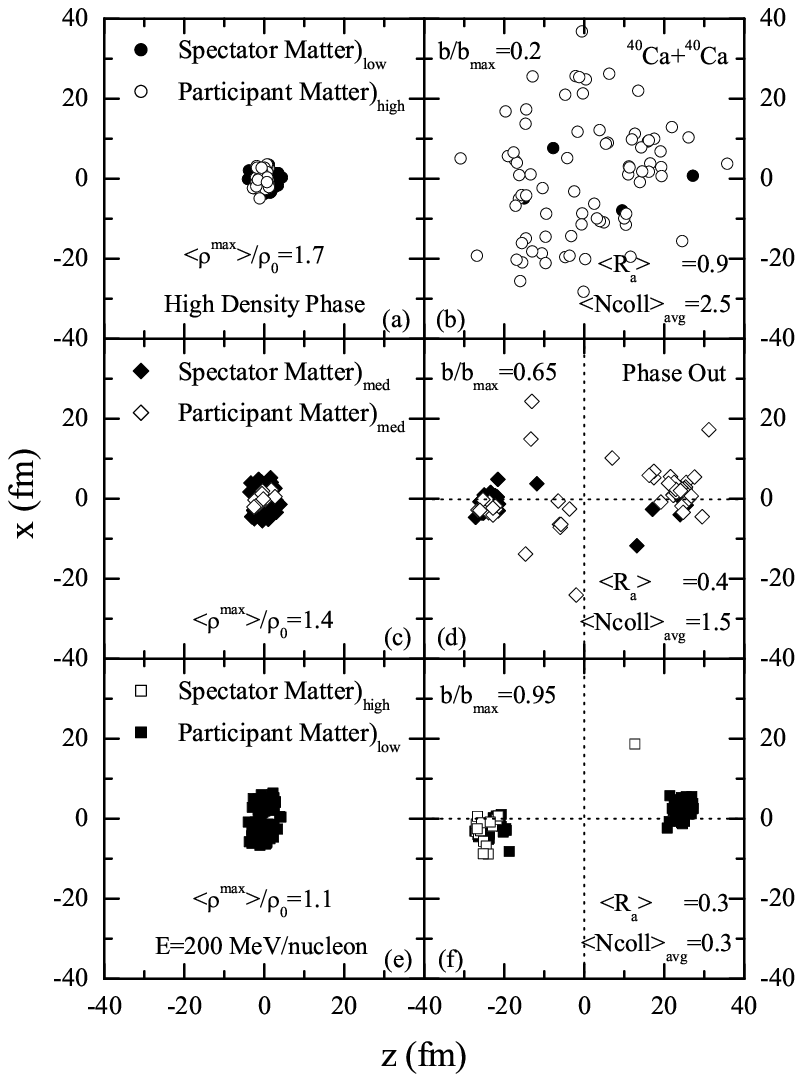}
\vskip -0.2 cm \caption{The time evolution of a single event of
$^{40}$Ca+$^{40}$Ca reaction at 200 MeV/nucleon. We use different
impact parameters $b/b_{max}=0.2$ (upper panel), 0.65 (middle
panel), and 0.95 (bottom panel). The solid (open) symbols, again,
represent the spectator (participant) matter. The high, low, and
med subscripts on the spectator/participant represent the
intensity of the matter, being high, low or medium for a
particular reaction. The displayed quantities like
$<\rho^{max}>/\rho_{0}$, $<Ncoll>_{avg}$, and $<$$R_{a}$$>$
(defined in Eq. (4)) are averaged over hundreds of
events.}\label{6.3p}
\end{figure}

\begin{figure}[!t] \centering
 \vskip 0.5cm
\includegraphics[angle=0,width=10cm]{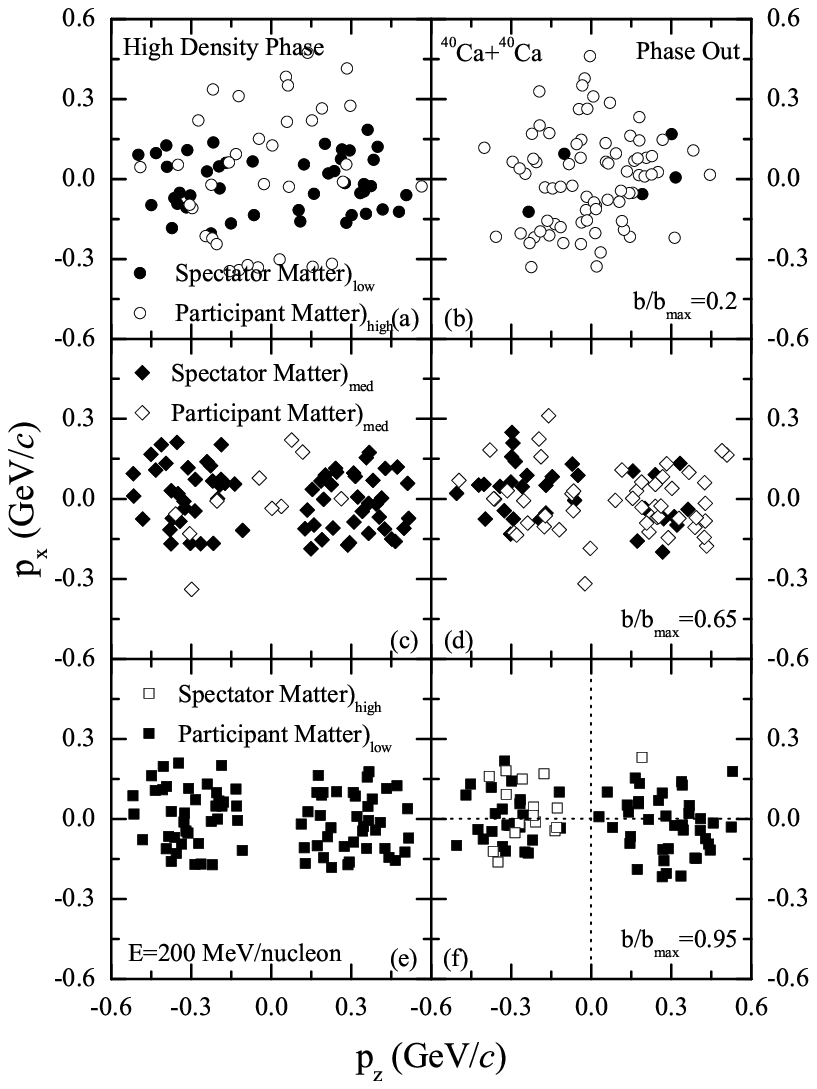}
\vskip -0.2 cm \caption{Same as Fig. \ref{6.3p}, but in momentum
space.}\label{6.4p}
\end{figure}

In Figs. \ref{6.3p} and \ref{6.4p}, we display the same evolution,
but at colliding geometries of $b/b_{max}=0.2$, 0.65, and 0.95,
respectively. The solid (open) symbols represent spectator
(participant) matter.
%%%%%%%%%%%%%%%%%%%%%%%%%%%%%%%%%%%%%%%%%%%%%%%%%%%%%%%%%%%%%%%%%%%%%%%%%%%%%%%%%%
The left panels of the figures are for the participant-spectator
matter formed at the time of high density. The right panels are
for the participant-spectator matter emerging at diluted phase of
the density. Note that at this state, the matter is cold and
fragmented. The reaction time for the high densities is around 20
fm/{\it c} for $b/b_{max}=0.2$, 0.65, and 0.95 as is also evident
from Fig. \ref{6.1p}. Note that the time of maximal density varies
linearly with the size of interacting nuclei.\cite{blat91,sood04}
From the figures, one notices a reduced intensity of the
participant matter with increase in the impact parameter. At
central impact parameters, nucleons are well scattered in the
momentum space. However, one can see two well separated momentum
spaces at higher impact parameters. This indicates that momentum
correlations remain preserved till the end of the reaction.
Further, central collisions lead to significant portion as
participant matter at high density. On the other hand, it is
almost spectator matter at peripheral geometries. One may say that
the reaction in central collisions is driven by the
nucleon-nucleon binary collisions. The mutual two and three-body
interactions dominate the physics at peripheral collision.
Interestingly, though, the high density phase is nearly the same
at $b/b_{max}=0.2$ and 0.65, the final stage evolutions is quite
different. In the formal one, it is more of complete disassembly
whereas it is a fire-ball picture in the later case. This
demonstration also points toward a very important clue for
multifragmentation of these nuclei. Almost complete participant
matter in central collisions does not allow any initial memory and
correlations to survive. Whereas a complete spectator matter at
the peripheral geometries preserves the initial correlations.
Therefore, in both these situations, one should not expect
intermediate mass fragments ($5\leq A\leq 65$) to be emitted. The
formation of the intermediate mass fragments needs both the
participant and spectator matter in a mild quantity. The
semicentral collisions are, therefore, prefect for the study of
emission of intermediate mass fragments. This has also been
reported as rise and fall of intermediate mass fragments in
several experimental and theoretical studies.\cite{tsan93,stoc89}
Similar rise and fall also occurs with the change in the incident
energies of the projectile.\cite{goss97} Interestingly, baring the
central collisions, no equilibrium is seen at semicental and
peripheral geometries. Further from the Fig. \ref{6.4p}, it is
also evident that no substantial changes occur after the maximal
high density in semicentral and peripheral collisions. However a
nearly complete equilibrium can be seen in central collisions.

%%%%%%%%%%%%%%%%%%%%%%%%%%%%%%%%%%%%%%%%%%%%%%%%%%%%%%%%%%%%%%%%%%%%%%%%%%%%%%%%%%
\begin{figure}[!t] \centering
 \vskip 0.5cm
\includegraphics[angle=0,width=10cm]{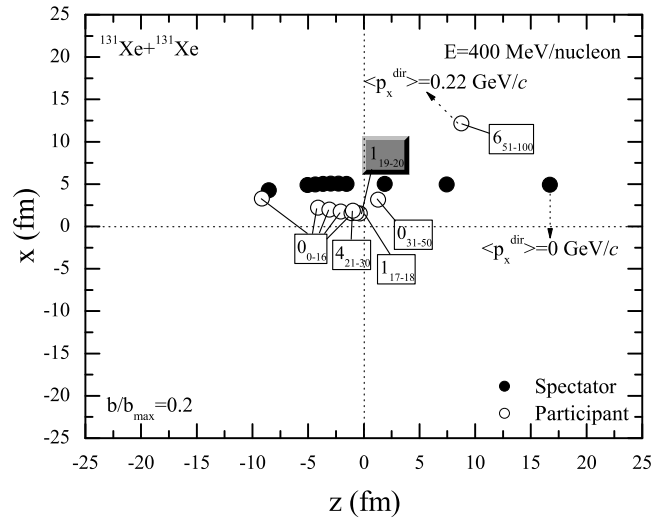}
\vskip -0.2 cm \caption{The trajectory of a single spectator
nucleon and participant nucleon for the collision of
$^{131}$Xe+$^{131}$Xe at incident energy of 400 MeV/nucleon. The
quantities in the boxes have superscripts and subscripts,
representing, respectively, the number of collisions and time
interval of these collisions. The shaded box is the interval for
maximal density.}\label{6.5p}
\end{figure}
\begin{figure}[!t] \centering
 \vskip 0.5cm
\includegraphics[angle=0,width=10cm]{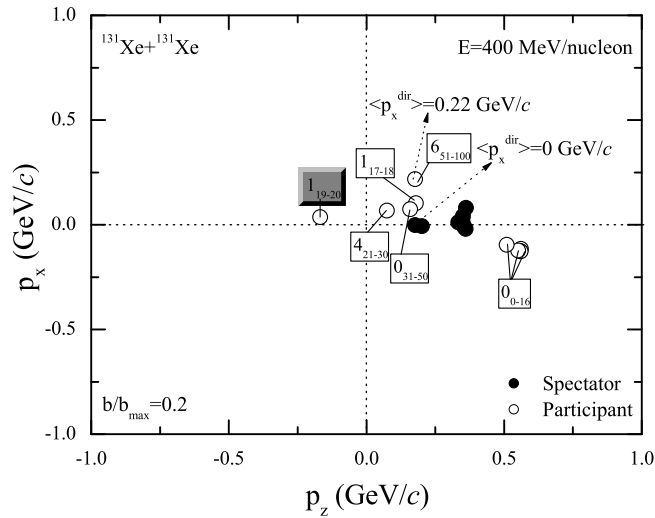}
\vskip -0.2 cm \caption{Same as Fig. \ref{6.5p}, but in momentum
space.}\label{6.6p}
\end{figure}
%%%%%%%%%%%%%%%%%%%%%%%%%%%%%%%%%%%%%%%%%%%%%%%%%%%%%%%%%%%%%%%%%%%%%%%%%%%%%%%%%%
Let us now look for the changes in the phase-space with reference
to the nucleon-nucleon collisions. In Figs. \ref{6.5p} and
\ref{6.6p}, we display the coordinate and momentum space,
respectively, for a single spectator nucleon (denoted by the solid
circle) and participant nucleon (denoted by the open circle). The
participant nucleon has suffered a large number of binary
collisions. The numeric values in the boxes represent the number
of collisions faced by a nucleon. The subscripts to these values
denote the time interval (in fm/{\it c}) during which these
collisions happened. From Fig. \ref{6.5p}, one notices that the
trajectory of spectator nucleon remains unchanged during the
entire duration of the collision. The participant nucleon,
however, shifts in the transverse direction after suffering
frequent collisions. Similarly, from Fig. \ref{6.6p}, momentum
space of the spectator nucleon remains unchanged. Whereas in case
of participant nucleon, \emph{z}-component of the velocity
decreases and \emph{x}-component of the velocity increases. The
interesting case is the velocity profile between 19-20 fm/{\it c}.
Here, velocity is negative, leading to the reversed
\emph{z}-coordinate between this interval. This time also happens
to be the time for the maximal density. We see that the
participant matter generates also the positive transverse
momentum.

\begin{figure}[!t] \centering
 \vskip 0.5cm
\includegraphics[angle=0,width=10cm]{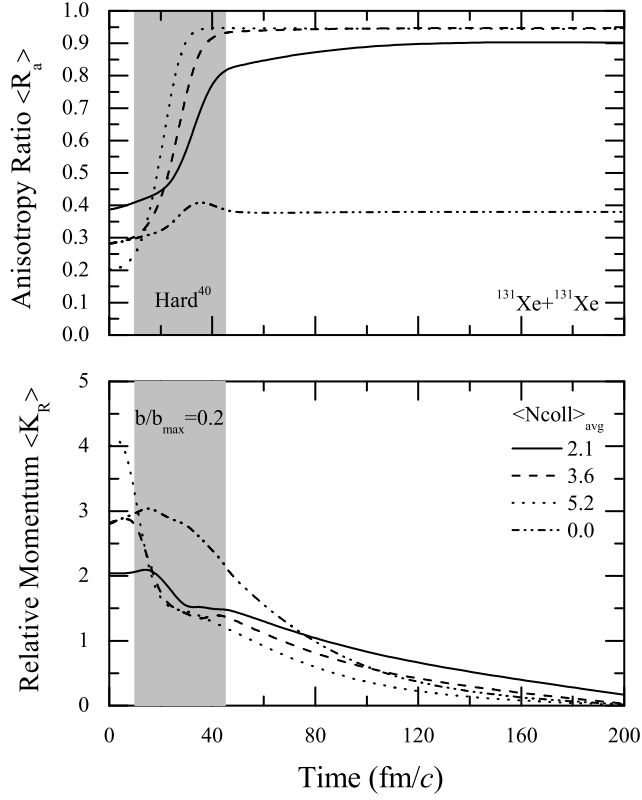}
\vskip -0.2 cm \caption{The anisotropy ratio $<$$R_{a}$$>$ and
relative momentum $<$$K_{R}$$>$ for the case of
$^{131}$Xe+$^{131}$Xe at $b/b_{max}=0.2$. Here different lines
represent the matter having different average collisions at
different incident energies. The shaded area represents the time
of high density $>\rho_{0}$ for different reactions displayed
here.}\label{6.7p}
\end{figure}

%%%%%%%%%%%%%%%%%%%%%%%%%%%%%%%%%%%%%%%%%%%%%%%%%%%%%%%%%%%%%%%%%%%%%%%%%%%%%%%%%%
In Fig. \ref{6.7p}, we display the time evolution of anisotropy
ratio $<$$R_{a}$$>$
and relative momentum $<$$K_{R}$$>$ for different average
collisions experienced in a reaction. The $<$$R_{a}$$>$ ratio is
defined as
\begin{equation}
<R_{a}> =
\frac{\sqrt{p_{x}^{2}}+\sqrt{p_{y}^{2}}}{2\sqrt{p_{z}^{2}}}.
\label{ar}
\end{equation}
\begin{figure}[!t] \centering
 \vskip 0.5cm
\includegraphics[angle=0,width=10cm]{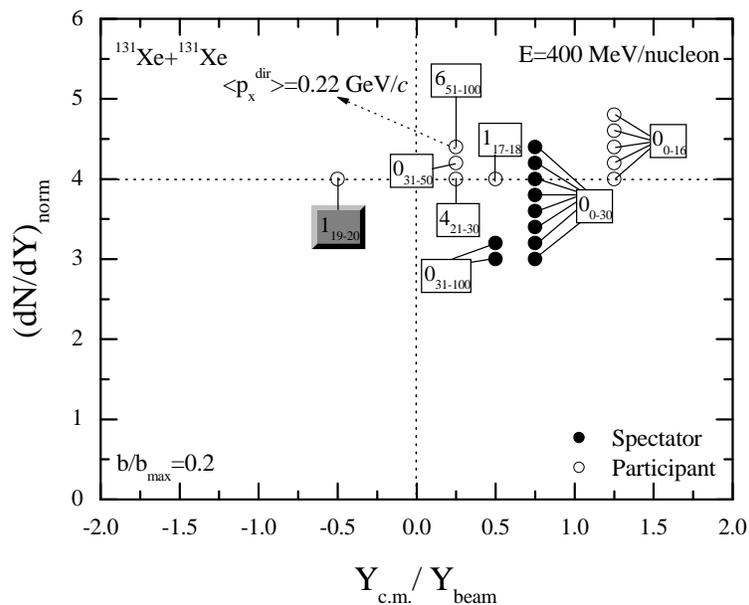}
\vskip -0.2 cm \caption{The normalized rapidity distribution
(dN/dY)$_{norm}$ as a function of Y$_{c.m.}$/Y$_{beam}$. Here a
single spectator nucleon and participant nucleon at different time
steps is taken. The shaded box is the interval for maximal
density.}\label{6.8p}
\end{figure}
This anisotropy ratio is an indicator of the global equilibrium of
the system. This represents the equilibrium of the whole system
and does not depend on the local positions. The full global
equilibrium averaged over large number of events will correspond
to $<$$R_{a}$$>$ = 1. The second quantity, namely the relative
momentum $<$$K_{R}$$>$ of two colliding Fermi spheres, is defined
as
\begin{equation}
<K_{R}> = <|\vec{P}_{P}(\vec{r},t)-\vec{P}_{T}(\vec{r},t)|/\hbar>,
 \label{kr}
\end{equation}
where
\begin{equation}
\vec{P_{i}}(\vec{r},t) =
\frac{\sum_{j=1}^{A}\vec{P_{j}}(t)\rho_{j}(\vec{r},t)}
{\rho_{j}(\vec{r},t)}~~~~~         i=1,2.
\end{equation}
Here $\vec{P_{j}}$ and $\rho_{j}$ are the momentum and density of
the \emph{j}th particle and \emph{i} stands for either projectile
or target. The $<$$K_{R}$$>$ is an indicator of the local
equilibrium because it depends also on the local position
\emph{r}. As expected, the anisotropy ratio increases or remains
almost constant whereas relative momentum decreases with increase
in the number of collisions indicating better degree of global
thermalization. The smaller value of ($<$$K_{R}$$>$) also
indicates toward the better thermalization of the matter. However,
once nucleon-nucleon collisions exceeds a certain limit, the
thermalization does not improve with further binary collisions.
Rather the nucleons prefer transparency in the system. Naturally,
the higher energy projectile has a lower value of ($<$$R_{a}$$>$)
at the start [conversely, the larger value of ($<$$K_{R}$$>$)].
Further, as is also evident, ($<$$R_{a}$$>$) ratio saturates as
soon as high density phase is over. In other words,
nucleon-nucleon collisions happening after high density phase do
not change the momentum space significantly. One also notices that
in the absence of nucleon-nucleon collisions (in Vlasov mode),
almost no change is seen in $<$$R_{a}$$>$.

%%%%%%%%%%%%%%%%%%%%%%%%%%%%%%%%%%%%%%%%%%%%%%%%%%%%%%%%%%%%%%%%%%%%%%%%%%%%%%%%%%
In Fig. \ref{6.8p}, we show the normalized rapidity distribution,
for a single spectator and participant nucleon. Although, the
dN/dY will have the same value for both these particles, we have
shifted the spectator and participant in \emph{y}-axis for better
clarity. Again, the values in the boxes represent the number of
collisions suffered by a nucleon. The subscripts to these values
represent the time interval (in fm/{\it c}). One notices that the
participant nucleon, after suffering significant binary
collisions, shifts from the projectile rapidity to midrapidity
region. On the other hand, the spectator nucleon remains in the
projectile rapidity region during the entire reaction.

\section{Summary}

We have presented the simulations of heavy-ion collisions in terms
of participant-spectator matter. We find that the
participant-spectator matter depends crucially on the collision
dynamics and on history of the nucleon. We see that the important
changes in the momentum space are due to the binary collisions
experienced during the phase of high density. This otherwise was
not possible with the mean field alone. The collisions push the
nucleons into midrapidity region that is responsible for the
formation of participant matter. This also indicates
thermalization in heavy-ion collisions. Interestingly, large
number of collisions, nevertheless, do not always guarantees a
better degree of thermalization of the reaction.

\end{document}